\documentclass[a4paper,fleqn,usenatbib]{mnras}

\usepackage{newtxtext,newtxmath}

\usepackage[T1]{fontenc}
\usepackage{ae,aecompl}
\pdfoutput=1
\usepackage{graphicx}	
\usepackage{amsmath}	
\usepackage{amssymb}	

\newcommand{\ubv}{\ensuremath{UB\,V}}
\newcommand{\logg}{\ensuremath{\log g}}
\newcommand{\msun}{\ensuremath{{\rm M}_\odot}}

\newcommand{\vsini}{\ensuremath{{\upsilon}\sin i}}
\newcommand{\kms}{\,km\,s$^{-1}$}

\title[Rotational properties of mCP stars]{An Investigation of the Rotational Properties of Magnetic Chemically Peculiar Stars}

\author[M. Netopil et al.]{
Martin Netopil,$^{1}$\thanks{E-mail: mnetopil@physics.muni.cz}
Ernst Paunzen,$^{1}$
Stefan H{\"u}mmerich,$^{2,3}$
Klaus Bernhard,$^{2,3}$
\\
$^{1}$Department of Theoretical Physics and Astrophysics, Masaryk University, Kotl\'a\v{r}sk\'a 2, 611 37 Brno, Czech Republic\\
$^{2}$American Association of Variable Star Observers (AAVSO), Cambridge, USA\\
$^{3}$Bundesdeutsche Arbeitsgemeinschaft f{\"u}r Ver{\"a}nderliche Sterne e.V. (BAV), Berlin, Germany\\
}

\date{Accepted 2017 March 15. Received 2017 March 15; in original form 2017 January 2}

\pubyear{2017}

\begin{document}
\label{firstpage}
\pagerange{\pageref{firstpage}--\pageref{lastpage}}
\maketitle

\begin{abstract}
The magnetic chemically peculiar (mCP) stars of the upper main sequence exhibit strong, globally organized magnetic fields that are inclined to the rotational axis and facilitate the development of surface abundance inhomogeneities resulting in photometric and spectroscopic variability. Therefore, mCP stars are perfectly suited for a direct measurement of the rotational period without the need for any additional calibrations. We have investigated the rotational properties of mCP stars based on an unprecedentedly large sample consisting of more than 500 objects with known rotational periods. Using precise parallaxes from the $Hipparcos$ and $Gaia$ satellite missions, well-established photometric calibrations and state-of-the-art evolutionary models, we have determined the location of our sample stars in the Hertzsprung-Russell diagram and derived astrophysical parameters such as stellar masses, effective temperature, radii, inclinations and critical rotational velocities. We have confirmed the conservation of angular momentum during the main sequence evolution; no signs of additional magnetic braking were found. The inclination angles of the rotational axes are randomly distributed, although an apparent excess of fast rotators with comparable inclination angles has been observed. We have found a rotation rate of $\upsilon/\upsilon_{\rm crit} \geq 0.5$ for several stars, whose characteristics cannot be explained by current models. For the first time, we have derived the relationship between mass and rotation rate of mCP stars, and provide an analysis that links mass and rotation with magnetic field strength. Our sample is unique and offers crucial input for forthcoming evolutionary models that include the effects of magnetic fields for upper main sequence stars.
\end{abstract}




\section{Introduction} \label{introduction}

Rotation is a key property of stars, which is correlated with many stellar properties and influences numerous stellar observables. It is a strong function of mass and evolutionary state, and can thus be used to derive relevant information on the investigated objects. Observed equatorial rotational velocities range from about a few m\,s$^{-1}$ up to $\sim$400\kms\ for the most massive stars \citep[e.g.][]{zorec12}, implying that some stars are rotating at a large fraction of their break-up velocity.

Among (single) main sequence (MS) stars, two rotational velocity regimes exist, with the transition occurring at a temperature of about 6200\,K (late F-type; mass $\sim$1.3\,\msun). This transition has come to be known as the Kraft break \citep{kraft67} and is connected to the presence (or absence) of a substantial surface convective zone that is needed to generate the magnetic winds that are believed to play an important role in angular momentum (AM) loss. Early-type stars hotter than $\sim$6200\,K are generally fast rotators, while the cooler, late-type stars tend to be slow rotators with typical periods on the order of tens of days \citep[e.g.][]{zorec12}.

However, an excess of slow rotators among early MS stars, particularly among the late B, A and early F stars, has long been established \citep[e.g.][]{slettebak54}. Rotational velocities in this region of the Hertzsprung-Russell diagram (HRD) roughly follow a bimodal distribution \citep[][]{abt95}, with a clear tendency of the slow rotators to deviate from normality in regard to chemical composition. Thus, most \citep[but not all, e.g.][]{wolff68,zorec12} of these abnormally slow rotators are so-called chemically peculiar (CP) stars, which have attracted much attention concerning the mechanisms behind the observed peculiarities, including the cause of the observed low rotational velocities.

About 10$-$15\% of upper MS stars between spectral types early B and early F are characterized by peculiar abundances of one or several elements. Accordingly, these objects are referred to as CP stars. The observed abundance anomalies are generally explained by the interplay of selective processes (radiative levitation, gravitational settling) operating in the calm radiative atmospheres of these stars \citep{richer00}.

CP stars are commonly divided into four subgroups \citep{preston74}: CP1 stars (the metallic-line or Am/Fm stars),  CP2 stars (the magnetic Bp/Ap stars), CP3 stars (the HgMn stars) and CP4 stars (the He-weak stars). Additional groups of CP stars have been defined, like e.g. the He-strong stars. CP2 and CP4 stars\footnote{For ease of use, CP2/4 stars are referred to hereafter as magnetic CP (mCP) stars.}, which are the subject of the present investigation, are set apart from other groups of CP stars by the presence of globally organized magnetic fields from about 300\,Gauss (G) to several tens of kiloGauss (kG) \citep[e.g.][]{auriere07,kochukhov11}.

The mCP stars are notorious for showing a non-uniform distribution of chemical elements on their surfaces, which results in the formation of spots and patches of enhanced element abundance \citep{michaud81}. Therefore, as a result of rotation, strictly periodic changes are observed in the spectra and brightness of many CP2 stars, which are well described by the oblique rotator model \citep{stibbs50}. CP2 stars exhibiting photometric variability are traditionally referred to as $\alpha^2$ Canum Venaticorum (ACV) variables.

\subsection{Rotation of magnetic CP stars}  \label{rotation_mCP}

During the past decades, many investigators have contributed to the present state of our knowledge of the rotational properties of CP stars. A thorough discussion and summary of these contributions is beyond the scope of this paper; however, a short overview over the field is given below, with special emphasis on the mCP stars.

It has been well established that CP stars rotate on average more slowly than normal stars of the same spectral types \citep[e.g.][]{sargent64,conti65,slettebak70,abt95,zorec12}, and that this discrepancy is not due to CP stars being rapid rotators seen pole-on \citep[e.g.][]{wolff67,preston70}. It has therefore been suggested that slow rotation (equatorial velocity $\la$\,120\kms; \citealt{murphy14}) is a necessary \citep[but not sufficient, e.g.][]{wolff68,zorec12} preliminary for the development of chemical peculiarities.

CP stars do not simply constitute a slowly rotating trail of ``normal'' A-type stars; instead, their rotational velocities follow a Maxwellian distribution with an average value 3$-$4 times lower \citep{preston70,wolff81,stepien98}. This bimodal distribution of rotational velocities among early MS stars is e.g. clearly illustrated in the work of \citet{abt95}.

In consequence, CP stars were assumed to form a separate population, for which a special mechanism of AM loss is operating or has been operating in the past. CP1 stars are mostly members of short-period binary systems (2\,$\leq$\,$P$\textsubscript{orb}\,$\leq$\,10 d), and it has been postulated that tidal interactions have reduced rotational velocities in this group of CP stars \citep{abt09}.

The situation is quite different for mCP stars, which are less often found in binary systems \citep[e.g.][]{gerbaldi85,north04,bernhard15a}. However, they are characterized by measurable magnetic fields that may attain strengths of several kG, which indicates a possible relation between the observed slow rotation and the presence of a strong magnetic field.

There exists a population of very slowly rotating objects among the mCP stars \citep[e.g.][]{wolff75}, with periods that must in some cases be of the order of $\sim$300 yr, and may possibly reach up to 1000 yr \citep{mathys16b}. Past studies have established an apparent excess of these extremely slow rotators, in which a very efficient braking mechanism seems to be operating. Within a given sample of mCP stars, the very long-period objects seem to be overrepresented by a factor of 40$-$60 \citep{preston70,wolff75}. However, before the background of the huge increase in the number of known rotation periods of mCP stars during the last decades, it would be very interesting to reinvestigate this matter. The actual excess factor, however, is not straightforward to determine and beyond the scope of the present investigation. 

Both short- and very long-period mCP stars share the same properties \citep{wolff75}, which suggests a common origin of the observed variability. Although there have been attempts to attribute the observed extremely long-period variations to some other cause than rotation \citep[magnetic cycles; precession, cf. e.g.][]{bychkov12}, it seems secure that the oblique rotator model is applicable throughout the whole period range of mCP stars and that the period of the observed variations, regardless of the involved time-scales, is the rotation period. This leads to a spread in rotational velocities of about 5$-$6 orders of magnitude among the A-type stars, which is unique among non-degenerate stars \citep{mathys15}.

The peculiar distribution of rotation periods among mCP stars presents a major theoretical challenge and has not been fully understood. Much attention has been devoted to the question of the origin of the observed abnormally slow rotational periods of mCP stars. The assumption that mCP stars are formed from proto-stellar clouds exhibiting very low AM could not be substantiated by observations of mCP stars belonging to clusters and associations \citep[cf.][]{stepien98}. The main focus of the discussion, therefore, has been on the question whether the observed slow rotation is acquired during life on the MS or whether AM is lost during the pre-main sequence (PMS) phase of stellar evolution \citep[cf. the discussion in][]{stepien00}.

Theoretical support and observational evidence were coming forth in favour of both approaches. Earlier work suggested magnetic braking to occur during the MS phase \citep{havnes71} and observational evidence seemed to back this up \citep{stift76,abt79,wolff81}. With the accumulation of more data, however, it was established that the period distribution of mCP stars in young open clusters is indistinguishable from that of much older field stars \citep{hartoog77,borra85,klochkova85}, thereby disproving the earlier assumption. It is worth mentioning, though, that a lengthening of the rotational period has been found in some mCP stars \citep{mikulasek09,townsend10}, which has been interpreted as braking due to magnetized winds.

\citet{north85} discussed the occurrence of rotational braking in mCP stars by investigating the correlation between age and periods of field mCP stars, using gravity as a proxy for stellar age. His results indicated an anticorrelation between the investigated parameters -- as log(g) decreases, rotational periods increase. This is entirely consistent with the conservation of AM during stellar evolution; no suggestion for the existence of a braking mechanism operating during the MS phase was found. Some years later, these results were verified using the newly available $Hipparcos$ data by the same author for mCP stars exhibiting enhanced abundances of Silicon \citep{north98}. These findings were backed up e.g. by \citet{stepien98}, who also did not find evidence for a significant AM loss on the MS and concluded that, in agreement with the general consensus, mCP stars must lose a significant amount of their initial AM in the PMS phase of evolution. More recently, a study of the PMS progenitors of the mCP stars \citep{alecian13} suggests that all the AM loss is indeed  completed at very early evolutionary stages.

In their thorough investigation into the evolutionary status of mCP stars, \citet{kochukhov06} studied the relation of rotation period and surface gravity and provided evidence that older stars rotate more slowly. It was found that, for the more massive mCP stars ($M$\,$\geq$\,3.0\,\msun), no evidence exists for significant changes in AM during the MS lifetime -- the observed correlation of period and age is fully explained by loss of inertia due to stellar evolution. The situation was less clear for the lower mass mCP stars. While no results were found for the mass regime $M$\,$\leq$\,2.0\,\msun, the authors do not rule out that mCP stars in the mass range 2.0\,\msun$\leq$\,$M$\,$\leq$\,3.0\,\msun\ might have undergone AM loss during the MS phase.

\section{Compilation of the sample}

New or improved data on the rotational periods of mCP stars have appeared recently in the literature, which allow the investigation of an unprecedentedly large sample of stars with known rotational periods. In the following, a short overview over the sources is given, from which rotational periods of our programme stars were drawn.

As a first step, the catalogue of mCP star periods compiled by \citet{renson01} and the International Variable Star Index of the AAVSO (VSX; \citealt{watson06}) were searched for data on rotational periods. The latter compilation incorporates the data of the General Catalogue of Variable Stars \citep{gcvs} and is the most up-to-date source on variable star data. In addition to that, periods were drawn from \citet{wraight12}, who analysed the light curves of a large sample of mCP stars using data obtained with the $STEREO$ spacecraft. About 80 ACV variables were identified in this way.

The All Sky Automated Survey (ASAS) is a project that aims at continuous photometric monitoring of the whole sky, with the ultimate goal of detecting and investigating any kind of photometric variability \citep{pojmanski02}. Employing data from the third phase of the ASAS project (ASAS-3), \citet{bernhard15a} and \citet{huemmerich16} investigated known or suspected mCP stars from the \citet{rm09} catalogue. In total, 673 bona-fide ACV variables and candidates were identified. Likewise, \citet{bernhard15b} used publicly available data from the SuperWASP survey, which resulted in the discovery of an additional 80 new ACV variables. These results were also incorporated into the present investigation.

In the case of objects with multiple entries in any of the aformentioned sources, we have adopted the most recent results. In this way, periods were compiled for more than 1300 stars in total.

As next step, we searched for parallax measurements in the $Hipparcos$ catalogue \citep{leeuwen07} and the recent Data Release 1 of the $Gaia$ satellite mission \citep{gaiamission,gaiadr1}. Parallaxes with an accuracy better than 25\,\% were retrieved for about 880 objects, and the more accurate parallax was adopted if measurements were available in both data sets. Furthermore, we queried for photometric data that allow an estimate of effective temperature (see Section \ref{sect:astroph_para}). The validity of the mCP classification of our programme stars was investigated by employing the MK spectral types listed by \citet{skiff14}, photometric peculiarity indices in the $\Delta a$ and $Geneva$ systems \citep[][]{paunzen05} and additional information such as measurements of the magnetic field strength. Objects whose status as mCP stars is uncertain were excluded. Likewise, we decided against introducing correction terms for known spectroscopic binary systems and dropped these objects from our sample. However, the lack of appropriate photometric data has been the main limiting factor responsible for reducing our final sample to 492 CP2 and 28 CP4 stars.

\begin{figure}
	\includegraphics[width=\columnwidth]{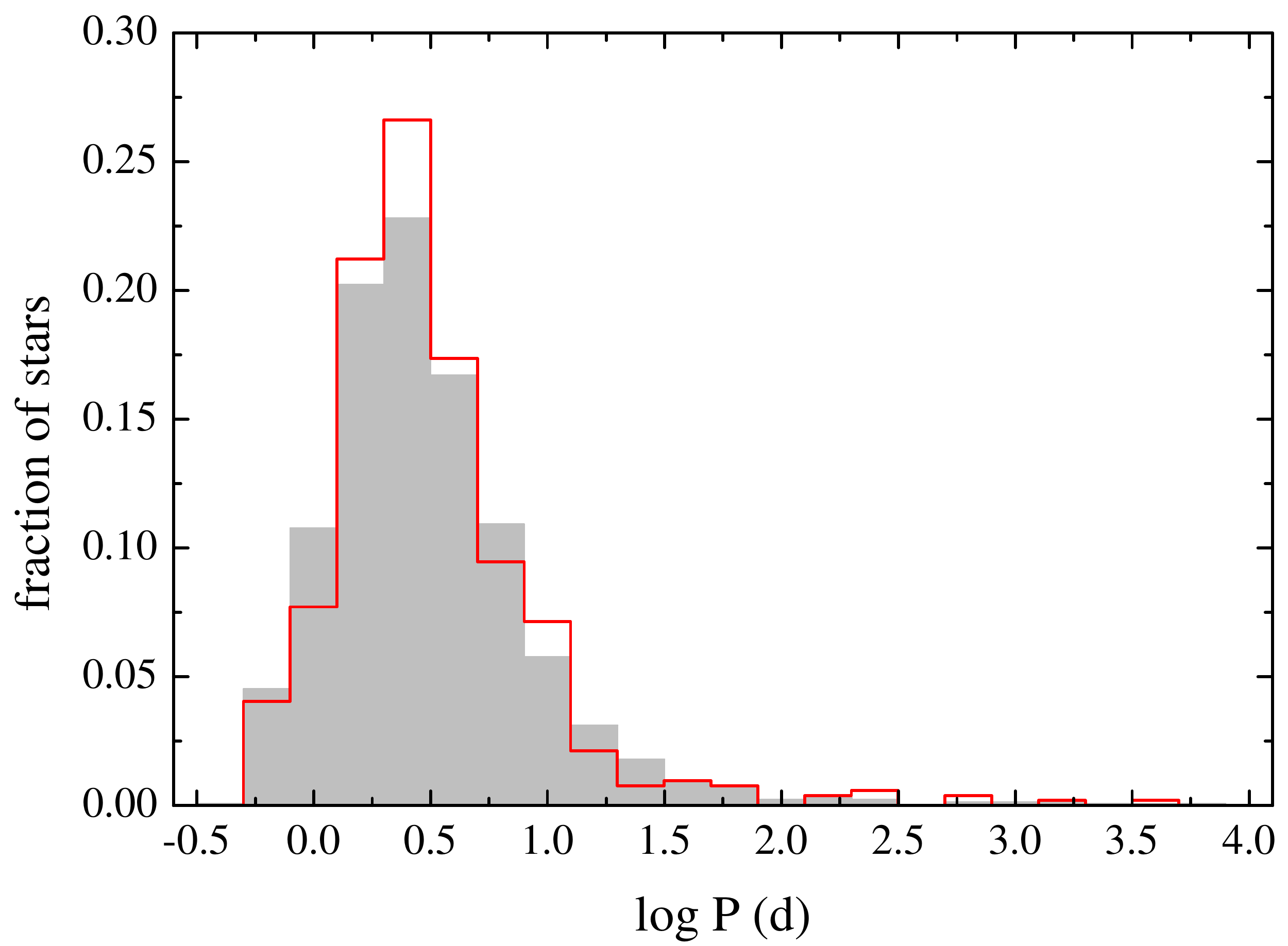}
    \caption{Distribution of rotational periods. The grey area represents the complete sample of about 1300 stars; the red line denotes the distribution in our final sample of 520 stars.}
    \label{fig:perioddistri}
\end{figure}

Table \ref{table_periods} provides an overview over the employed period sources and the number of period values for the stars that entered the final sample. Figure \ref{fig:perioddistri} shows the distribution of rotational periods in the original starting sample of about 1300 stars and in the final one. Both agree well, assuring that the selection process does not introduce a significant bias in the distribution.

\begin{table}
\caption{Period sources and number of period values taken from each source.}
\label{table_periods}
\begin{center}
\begin{tabular}{ll}
\hline 
Source & \#period values \\
\hline
Period compilation \citep{renson01} & 201 \\
VSX \citep{watson06} & 67 \\
STEREO \citep{wraight12} & 51 \\
ASAS-3 \citep{bernhard15a} & 98 \\
WASP \citep{bernhard15b} & 11 \\
ASAS-3 \citep{huemmerich16} & 92 \\
\hline
Number of stars in the sample & 520 \\
\end{tabular}
\end{center}
\end{table}

\section{Determination of astrophysical parameters}
\label{sect:astroph_para}
Spectroscopic effective temperatures are only available for a small number of mCP stars \citep{netopil08}, thus we had to fall back on photometric data to derive this parameter. We queried `The General Catalogue of Photometric Data' (GCPD\footnote{http://gcpd.physics.muni.cz/}) and \citet{paunzen15} for measurements in three photometric systems (\ubv, Str\"omgren--Crawford $uvbyH\beta$, and $Geneva$) which allow the calibration of effective temperature for mCP stars \citep{netopil08}. For B-type stars, the majority of objects in our sample, all systems provide an estimate of interstellar reddening \citep[see overview by][]{netopil08}. For cooler stars, however, we had to rely on reddening calibrations in the $uvbyH\beta$ system \citep{napi93}. Additionally, we adopted $BV$ photometric data from \citet{kharch01} and $JHK_{s}$ photometry from the Two Micron All Sky Survey \citep[][]{2mass}, as well as photometric data from the Wide-field Infrared Survey Explorer \citep[][]{wise}. Only measurements with S/N\,$>$\,5 were employed from the last two sources.

For the stars with an estimate of reddening, we also fitted the spectral energy distribution (SED), which was compiled from all available data, using the VO Sed Analyzer \citep[VOSA;][]{vosa} tool. \citet{wraight12} derived a temperature correction for CP2 stars based on SED fitting, using, however, a different SED fitting application. In order to investigate the validity of our approach, we have compared our results with those of the aforementioned study.  While for CP2 stars cooler than $T_{\rm SED} = 10000$\,K no correction is necessary, hotter stars in this group follow the linear relation $T_{\rm eff} (K) = 1110(240) + 0.889(20)T_{\rm SED}$, in good agreement with \citet{wraight12}. This result has been based on almost 300 objects for which we applied at least two temperature calibrations of \citet{netopil08}. Our sample of CP4 stars is considerably smaller; however, using 25 objects, we notice that $T_{\rm SED}$ is underestimated by 660 $\pm$ 470\,K.

Finally, we derived mean $T_{\rm eff}$ values, which have been based on at least two different calibrations. Most effective temperature values rely on four or more individual estimates. The mean standard deviation is 260 and 370\,K for CP2 and CP4 stars, respectively, but we associate uncertainties of at least 500 and 700\,K, as proposed by \citet{netopil08}. From this reference, we also adopt the bolometric correction for mCP stars as a function of temperature, in order to finally derive luminosities employing the $Hipparcos$ or $Gaia$ parallax, the reddening values, the compiled $V$ magnitudes, and the absolute bolometric magnitude of the Sun (4.74\,mag). The error of the luminosity takes into account an error propagation that includes fixed uncertainties of 0.05\,mag for the $V$ magnitude and $E(B-V)$, and the individual distance error. The bolometric correction was calculated as a function of effective temperature, thus the error of temperature contributes to the total luminosity uncertainty as well. Additionally, we adopt an error of 0.05\,mag for the bolometric correction calibration itself.

Several stellar evolutionary models are available in the literature, but the use of different input physics renders an appropriate choice complex \citep[see e.g. the comparison by][]{stancliffe16}. In the case of mCP stars, an additional error source is introduced by magnetic fields as missing input physics. Furthermore, mCP stars exhibit peculiar abundance patterns on the stellar surface, but the chemical bulk composition remains unknown. We have thus assumed solar metallicity -- an inference, which has been generally justified through the comparison of mCP cluster stars \citep{landstreet07} with the overall metallicity of their respective host clusters \citep{netopil13,netopil16}. Furthermore, all objects are reasonably close ($\ll$ 1\,kpc), so that solar metallicity is likely to be a reasonable assumption. However, we note that mCP stars have been proved to also exist in low-metallicity environments, such as the Large Magellanic Cloud \citep[][]{paunzen06,paunzen11}.

\begin{figure}
	\includegraphics[width=\columnwidth]{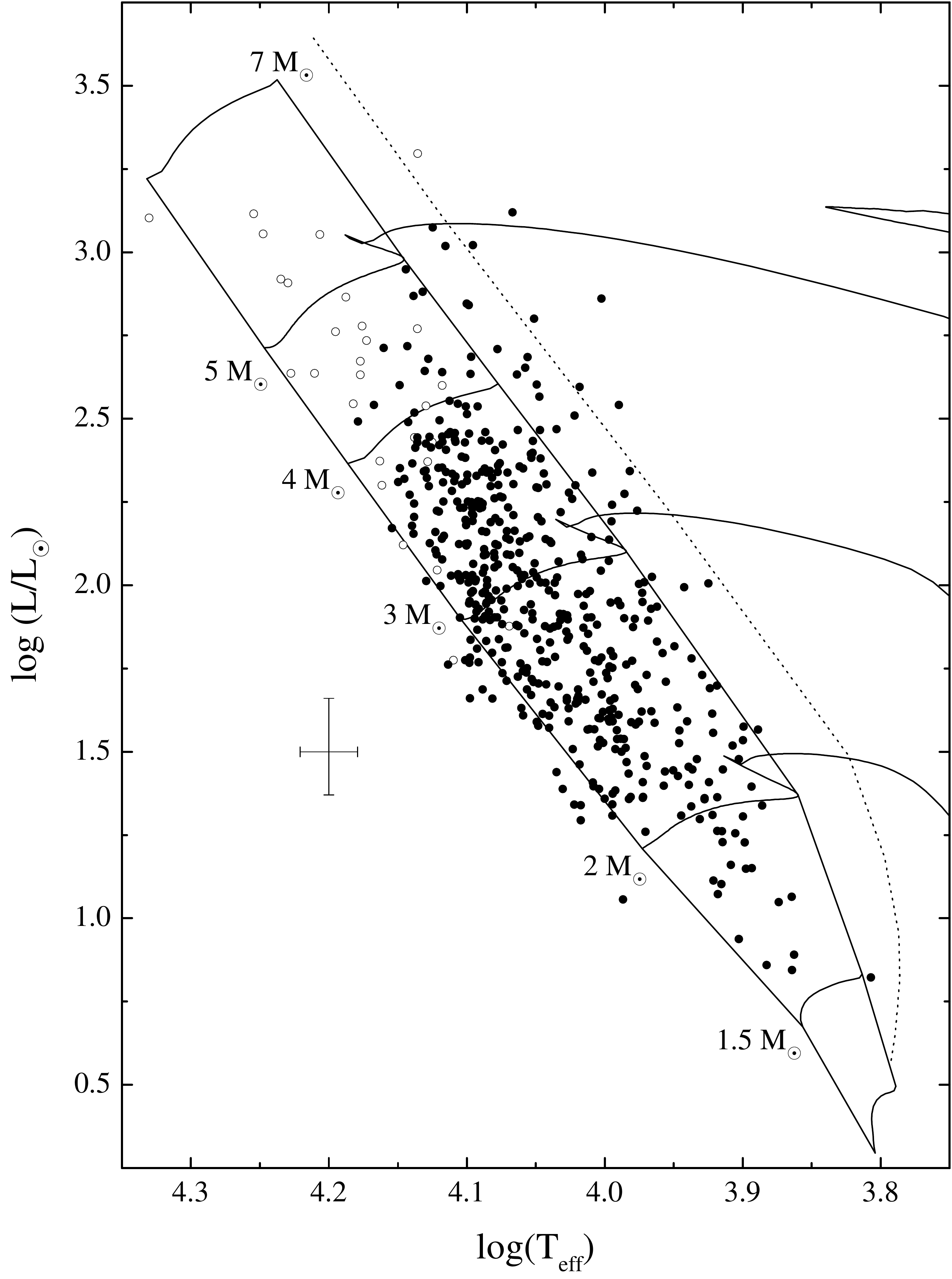}
    \caption{Positions of the final sample of mCP stars in the HRD. CP2 and CP4 stars are denoted by filled and open circles, respectively. Also shown are stellar evolutionary models by \citet{ekstrom12} and the terminal-age MS by \citet{bressan12} (dotted line). Some mass tracks and the typical (mean) error of the parameters are indicated. We note that two objects were excluded from the initial sample as discussed in Section \ref{sect:inclination}; therefore, the plot has been based on 518 objects.}
    \label{fig:hrd}
\end{figure}

Fig. \ref{fig:hrd} shows the positions of our programme stars in the HRD. We have finally used the evolutionary tracks for Z=0.014 of the Geneva group \citep{ekstrom12,georgy13}  because these models are also available for several rotation rates and are thus best suited to our purpose. However, for deriving stellar mass, we have adopted non-rotating models, because slow rotation is a well-known characteristic of mCP stars. We note that rotation does not have a significant impact on the derived mass; at a rotation rate of 40\,\% of the critical velocity, the derived masses are on average lower by only about 1\,\%, which is well below the estimated errors of the mass ($\la$ 15\,\%).

\begin{figure}
	\includegraphics[width=\columnwidth]{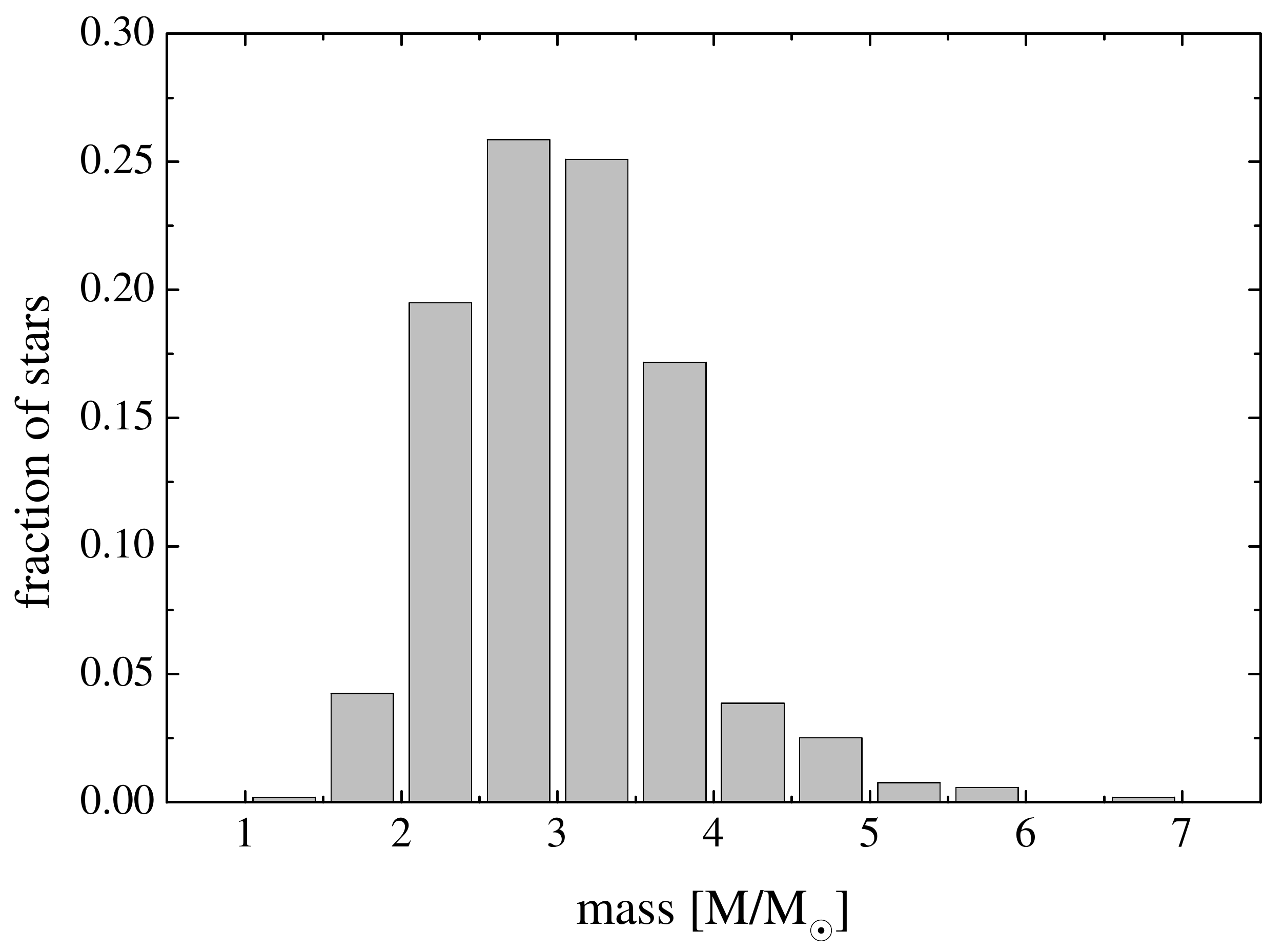}
    \caption{Mass distribution of the final sample of mCP stars as shown in Fig. \ref{fig:hrd}.}
    \label{fig:mass_histo}
\end{figure}

Fig. \ref{fig:mass_histo} illustrates the mass distribution of our sample stars. As further discussed in Section \ref{sect:inclination}, two objects were excluded, so that the plot has been based on 518 stars. From Figs. \ref{fig:hrd} and \ref{fig:mass_histo}, we can immediately infer that the programme stars cover the mass range from about 1.5 to almost 7\,\msun, and that most objects are encountered in the range 2.0\,\msun$\leq$\,$M$\,$\leq$\,4.0\,\msun. Some objects are located below the zero-age main sequence (ZAMS). However, taking error estimates into account, only few significant outliers remain. These could be due either to an effect of metallicity or the result of erroneous parallax measurements. For some stars, we have noticed significant differences between the $Hipparcos$ and $Gaia$ parallaxes, which might explain these objects' deviant positions off the MS. We note that masses for the stars below the ZAMS were derived by adopting the corresponding nearest positions on the ZAMS.

Similar effects might be responsible for the position of the stars in our sample which are located above the TAMS. In general, mCP objects are MS stars, but it remains unknown whether and to what extent photometrically visible spots are still present during the evolution off the MS. It has to be noted that, beside the effect of rotation, the width of the MS is also a matter of the adopted overshooting parameter in the models. \citet{ekstrom12} used $d_{\rm ov} = 0.10 H_{\rm P}$\footnote{$H_{\rm P}$ is the pressure scale height.} for $M >$ 1.7\msun, while for example \citet{bressan12} adopt $d_{\rm ov} \sim 0.25 H_{\rm P}$ for $M >$ 2\msun. We show the TAMS of the latter model for Z=0.014 in Fig. \ref{fig:hrd} and note that the position of the ZAMS agrees very well in both models. Recently, \citet{claret16} derived the overshooting parameter in a semi-empirical way and obtained $d_{\rm ov} \sim 0.20 H_{\rm P}$ for stars more massive than 2M$_{\odot}$, in reasonable agreement with the value adopted by \citet{bressan12}. There are only two objects (HD~40394 and HD~171247) that are located well above the MS band defined by \citet{bressan12}, and we have not found any hints for assuming a possible shift towards higher luminosities (e.g. caused by binarity).

The stellar evolution on the upper MS \citep[][]{Salaris2006} is defined by the fact that upper MS objects have fully mixed cores because of convection. Therefore, the abundance of hydrogen decreases uniformly throughout the core. In the final stages of its MS lifetime, the star contracts in an attempt to maintain energy production by increasing its core temperature. As hydrogen is exhausted, the star reaches the TAMS and establishes a hydrogen-burning shell source around a helium core (luminosity classes IV and III), followed by the development of a degenerate carbon--oxygen core (luminosity classes III and II).

On the basis of $Hipparcos$ data, \citet[][]{Paunzen1999} showed that the controversial status of luminosity class IV \citep[][]{Keenan1985} is apparent. There seems to be no distinction in absolute luminosity between stars of spectroscopic luminosity classes IV and V. Furthermore, a considerable overlap between these classes and luminosity class III exists, with only the brightest class III objects being well separated from those of class V. This implies that luminosity class IV should be rejected.

Do CP B- and A-type stars of luminosity class III or II exist? \citet[][]{Loden1989} examined 23 stars (four in binary systems) that had been classified as giant CP stars in the literature but found no evidence that these objects are considerably more luminous than MS stars of the same effective temperature. We have investigated the available spectral classifications and literature data for these objects. Most classifications are based on photographic plates from the Michigan spectral catalogues and indicate strong silicon lines. Since these lines are also enhanced in giants and employed as luminosity criterion, a corresponding misclassification is suggested, as also discussed in \citet[][]{Loden1989}. This has also been supported by CCD spectra, which became available later for some of these objects (e.g. HD 41089, HD 44293 and HD 152273) and show no peculiar composition. Luminosities from the $Hipparcos$ catalogue further delimitate the sample. To sum up, no good candidate for an evolved classical CP star remains. If such objects really exist, they seem to be very rare.

\citet[][]{Auriere2008} concluded that the slowly rotating (period of about 300\,d), active, single G-type giant EK Eri (HD 27536) is the descendant of a strongly magnetic MS CP star and inferred a mean surface magnetic field of about 270\,G. The observed activity and magnetic field strength are incompatible with the classical solar-dynamo 
requirements. Following the scenario evoked by the above mentioned authors, EK Eri offers the first direct evidence that CP2 stars evolve to late-type giants maintaining a significant stellar magnetic field. However, the questions to what extent stellar magnetic fields are stable and diffusion is functioning during the sub- and giant branch evolution remain largely unanswered.

\section{Rotation rates and inclination of rotational axes}
\label{sect:inclination}
\begin{figure}
	\includegraphics[width=\columnwidth]{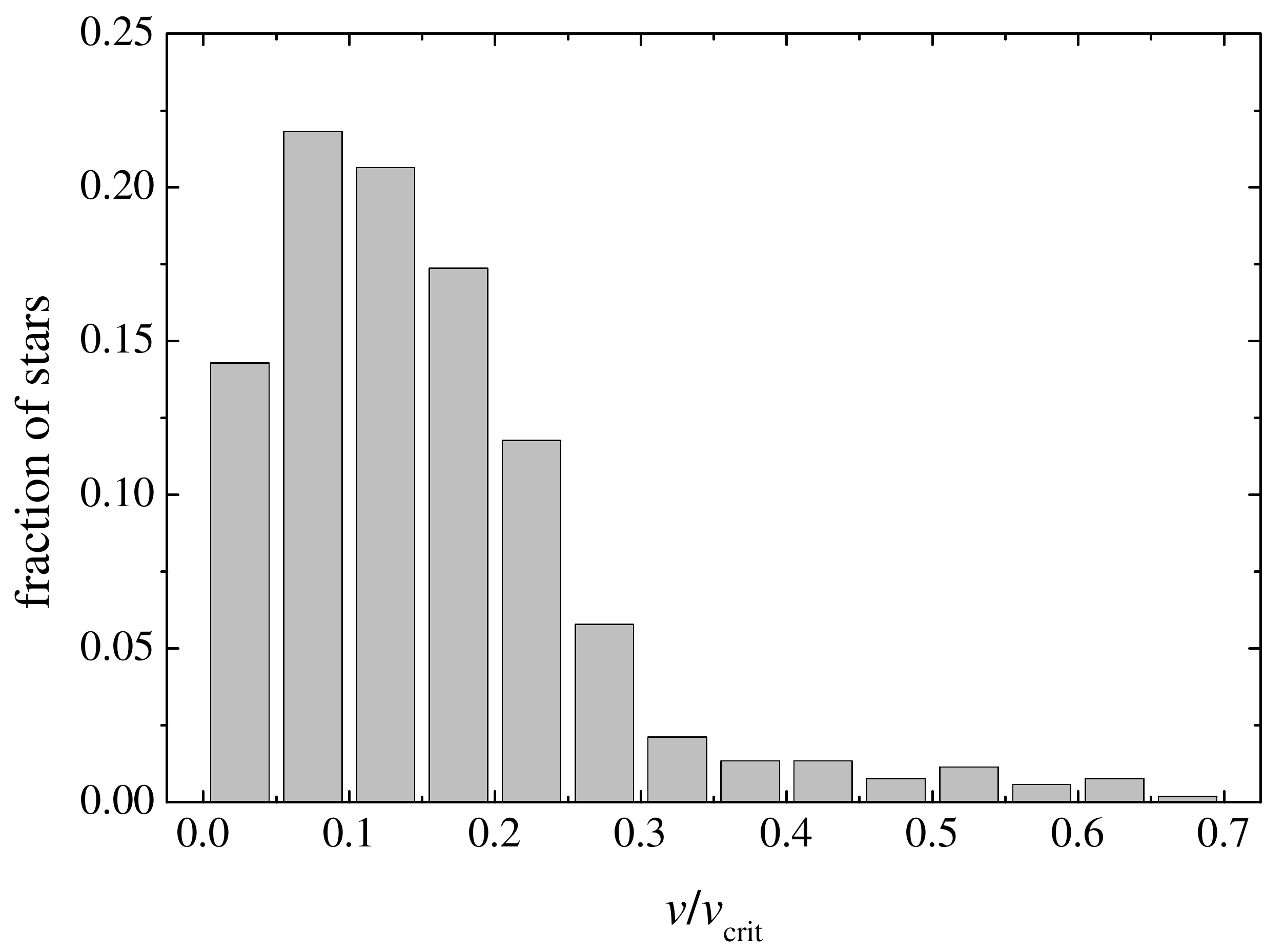}
    \caption{Distribution of the velocity ratio $\upsilon/\upsilon_{\rm crit}$ based on 518 mCP stars. Two objects were excluded from the initial sample as discussed in Section \ref{sect:inclination}.}
    \label{fig:vvcrit_distri}
\end{figure}

The compiled data allowed us to estimate stellar radii and to derive equatorial velocities, using the following equation. 
\begin{equation}
{V}_{\rm eq} ({\rm km\,s^{-1}})  = 50.579 R({\rm R}_\odot)/P(d)
\end{equation}
Furthermore, we used the first critical velocity as adopted e.g. by \citet{georgy13} to obtain the velocity ratio $\upsilon/\upsilon_{\rm crit}$ (see Fig. \ref{fig:vvcrit_distri}). Two objects (HD~162651 and HD~190576) show a physically implausible velocity ratio $\upsilon/\upsilon_{\rm crit}>1.2$, suggesting that the adopted periods probably do not represent the true rotational period, which is supported by a subsequent analysis of the available photometrical data. We thus excluded these objects.

Fig. \ref{fig:vvcrit_distri} shows that a small fraction of stars (14 objects) are fast rotators ($\upsilon/\upsilon_{\rm crit} \geq 0.5$) with an upper limit around 0.6, corresponding to ${V}_{\rm eq} \sim 200$\,\kms. This provides a stringent constraint for models of mCP stars, the efficiency of diffusion and its interplay with mixing processes. Furthermore, the question arises under what circumstances the magnetic field remains stable at these high rotational velocities. Only for the `slowest' rotating star in this group (HD~19832), an abundance analysis is available by \citet{shulyak10}, which indicates a typical mCP star abundance pattern and a dipolar magnetic field strength of about 1.5\,kG. In total, five fast rotators have a measured magnetic field strength of this order. We finally note that all these stars have comparable masses ($\sim$\,3\,\msun) and rotate more than twice as fast as the most massive mCP (CP4) stars.

Projected rotational velocities (\vsini) help us to estimate the inclination of the rotational axes, but also allow a check of the compiled parameters (period and stellar radius). We therefore queried for such data in the homogenized catalogue by \citet{glebocki05} and the additional literature, and retrieved measurements for 220 stars of our sample. For 25 objects, the data by \citet{levato96} indicate only an upper limit of 30\kms; therefore, these results were used for a plausibility control only.

There are 15 stars that show large errors or a \vsini/$V_{\rm eq}$ ratio $>$1, even after taking 2$\sigma$ into account; for these objects, no constraint on inclination angle could be derived. We note that we follow the standard error propagation for $\sigma(V_{\rm eq})$ and adopt 10\kms\ as a standard error for \vsini. The above mentioned outliers may be due to erroneous radii, periods, but also \vsini\ values -- unfortunately, for most stars, only one study covering the last parameter is available. A large fraction of these objects is made up of slow rotators (two-thirds with $\upsilon/\upsilon_{\rm crit} < 0.05$). Therefore, an exclusion of these stars might result in a distortion of the sample. We have thus chosen to keep them in the sample, but do not consider these objects in the analysis of the inclination distribution. For another 37 objects with $\sin i >1$, large inclination angles are most likely within the errors. The error of $i$ increases towards larger angles (but also towards smaller rotational velocities); we thus analyse the distribution of the rotational axes in the $\sin i$ domain.

\begin{figure}
	\includegraphics[width=\columnwidth]{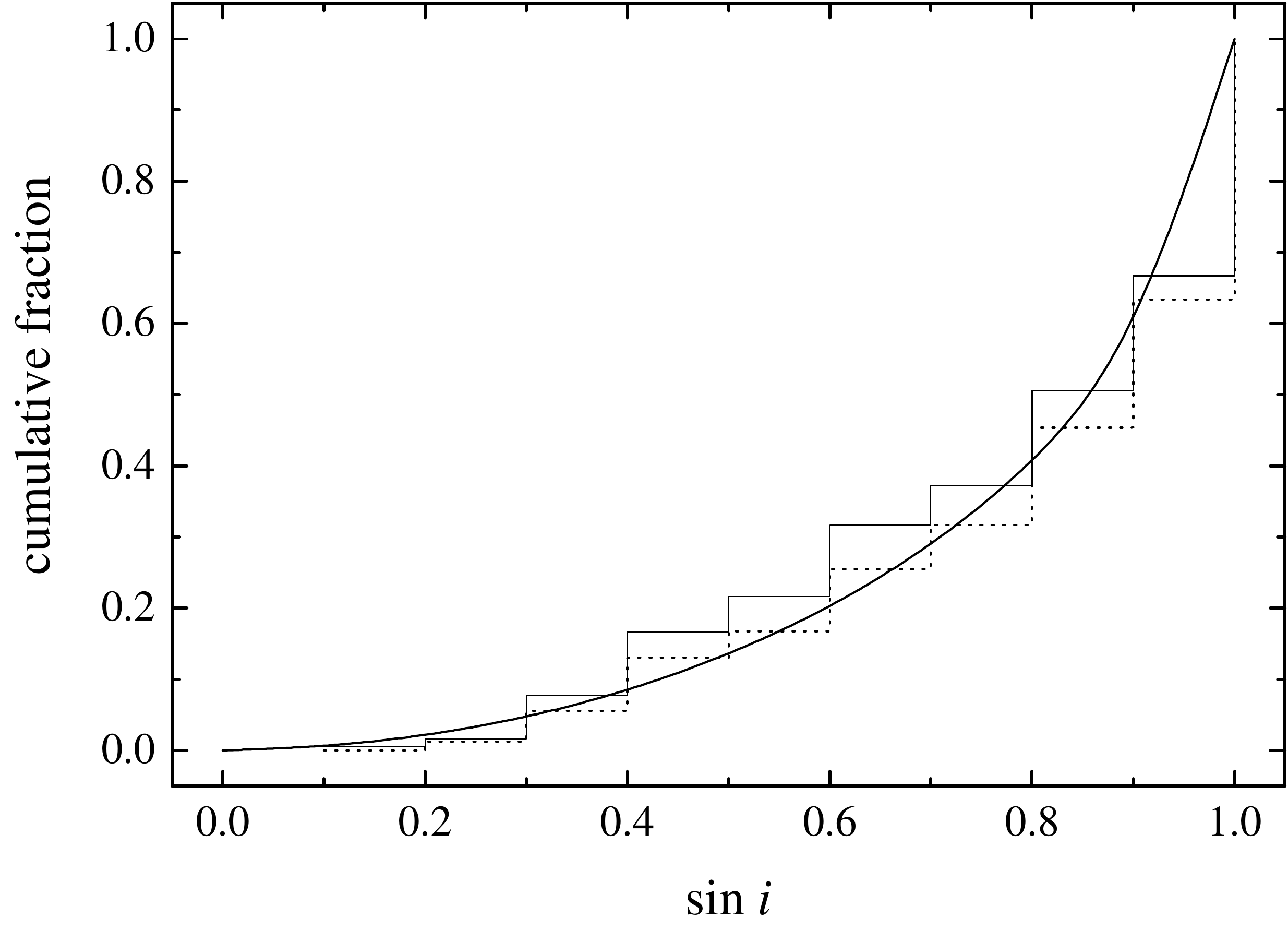}
    \caption{Cumulative distribution of observed $\sin\,i$ values of 180 stars (solid step line) and the theoretical expectation if the stellar rotation axes are randomly distributed (smooth solid line). The 37 objects with $\sin i >1$ that suggest within the errors a high inclination angle were added to the last bin. The dotted step line shows the distribution of the stars after the exclusion of 19 fast rotators ($\upsilon/\upsilon_{\rm crit} > 0.35$).}
    \label{fig:incli_distri}
\end{figure}

Fig. \ref{fig:incli_distri} has been based on 180 stars and confirms in principle previous conclusions \citep[e.g.][]{abt01} that rotational axes are randomly distributed. However, there is an excess of stars with moderate $i$ values ($\sim$\,20--40\degr), which might be caused by the presence of faster rotators ($\upsilon/\upsilon_{\rm crit} > 0.35$) in the sample. Their exclusion results in better agreement with the theoretically expected random distribution in this angle range, but also leads to a more pronounced deficit of stars seen almost pole-on. Certainly, the detection probability of the rotation period due to photometric variability caused by abundance spots decreases towards smaller inclination angles. However, as for the slow rotators, a cut at the fast rotator tail introduces a statistical bias as well, in particular because there are some well-confirmed faster rotators among mCP stars. Nevertheless, the reason for the concentration of fast rotators at lower inclination angles remains unclear. This might be owed to systematic errors in the determination of larger \vsini\ values and the neglection of the observed difference of effective temperature between the pole and the equator, to mention two possible causes.

\section{Evolution of rotational periods}

The stellar age or the fractional MS age are both quite model dependent owing to the adopted overshooting parameter and the resulting width of the MS. We have therefore resorted to using \logg\ as evolutionary parameter. A straightforward estimate is possible with
\begin{equation}
\log(g/g_\odot) = {\rm log}({M}/\msun)+4{\rm log}(T_{\rm eff}/T_{\rm eff \odot})-{\rm log}(L/{\rm L}_\odot)
\end{equation}
and the recommended IAU values for the Sun (\logg$_\odot$ = 4.438 and T$_{\rm eff \odot} $ = 5772\,K). It becomes obvious that the uncertainty of $T_{\rm eff}$ contributes most to the error of \logg\ (the typical error amounts to $\la$ 0.2\,dex). 

There certainly exists a dependence of rotation with mass; we have therefore grouped the stars according to their mass. The large number of mCP stars allowed us to define narrower mass groups than used in previous works, and the sample was split into groups with a bin size of 0.5\,\msun\ in the main mass range 2.0\,\msun$\leq$\,$M$\,$\leq$\,4.0\,\msun. We first investigated the dependence of rotational speed on mass. The periods, and -- to a somewhat smaller extent -- also $V_{\rm eq}$, show evolutionary changes. On the other hand, the $\upsilon/\upsilon_{\rm crit}$ ratio is almost constant for rigid rotators on the MS, in particular at slow rotation rates \citep[see e.g.][]{zorec12}.  

\begin{table}
	\centering
	\caption{Rotational properties of mCP stars. Indicated are the median values of the mass, the velocity ratio $\upsilon/\upsilon_{\rm crit}$, the observed velocity $V_{\rm eq, obs}$ and the velocity that we derived for the ZAMS using stellar evolutionary models $V_{\rm eq, ZAMS}$. The number of stars used for the calculation of $\upsilon/\upsilon_{\rm crit}$ and $V_{\rm eq, obs}$ is provided in the last column. The median absolute deviations are given in parentheses.}
	\label{tab:mean_velo}
	\begin{tabular}{lcccr} 
		\hline
		Mass  & $\upsilon/\upsilon_{\rm crit}$ & $V_{\rm eq, obs}$  & $V_{\rm eq, ZAMS}$ & No. of stars \\
		
		(\msun)  &  & (\kms) & (\kms) & \\

		\hline
		1.83(12) & 0.08(3) & 28(12) & 31 & 21/18\\
		2.31(11) & 0.10(4) & 37(15) & 42 & 80/63\\
		2.76(11) & 0.12(5) & 49(16) & 54 & 125/94\\
		3.22(13) & 0.14(6) & 55(21) & 63 & 110/79\\
		3.70(12) & 0.15(6) & 62(23) & 73 & 81/37\\
		4.47(27) & 0.17(5) & 85(16) & 87 & 37/15\\
		\hline
	\end{tabular}	
	
\end{table}

\begin{figure}
	\includegraphics[width=\columnwidth]{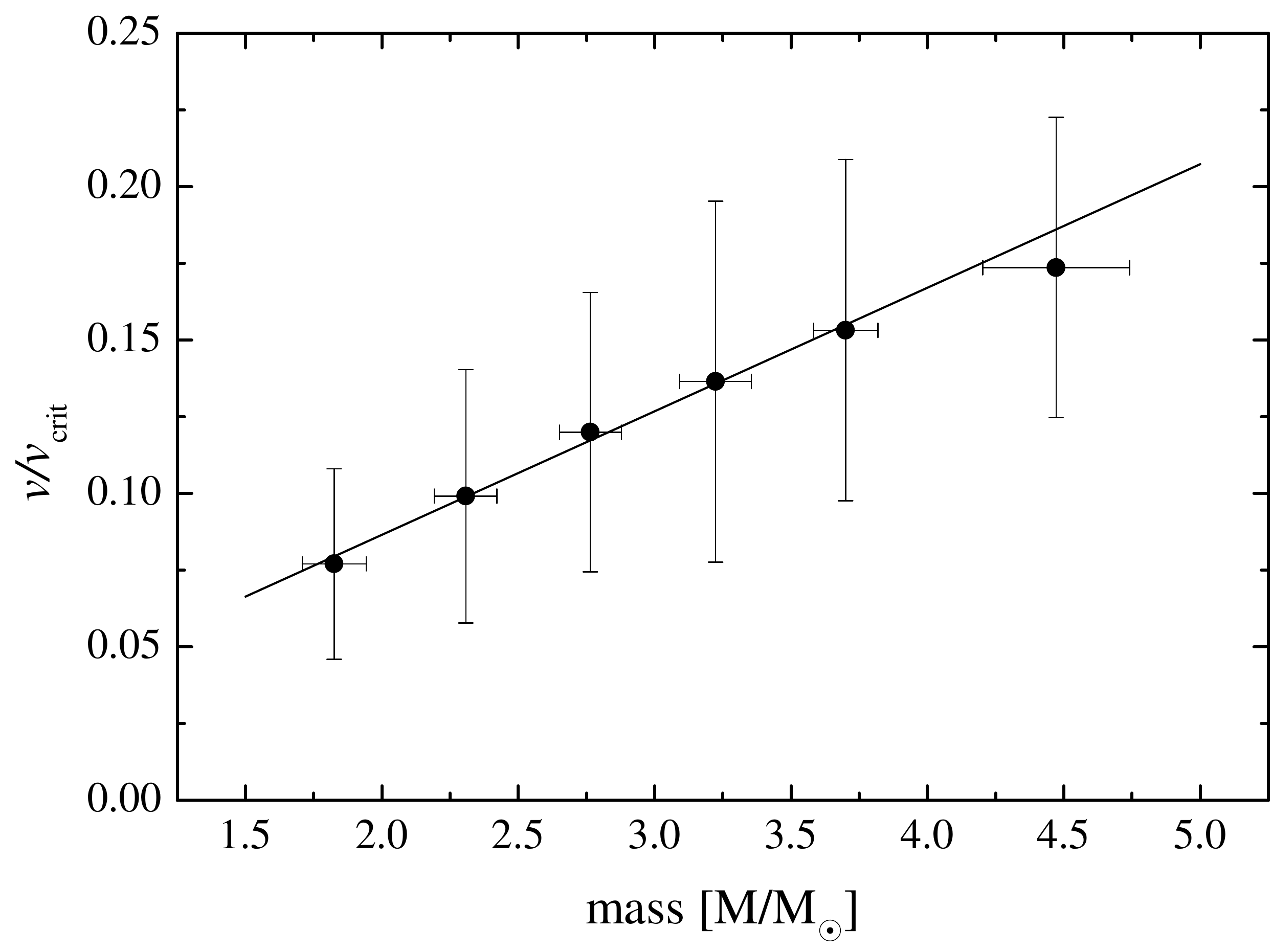}
    \caption{The velocity ratio $\upsilon/\upsilon_{\rm crit}$ as a function of mass as listed in Table \ref{tab:mean_velo}. The solid line is the linear dependence according to Equation \ref{vcritmass}.}
    \label{fig:mass_velo}
\end{figure}

\begin{figure*}
	\includegraphics[width=170mm]{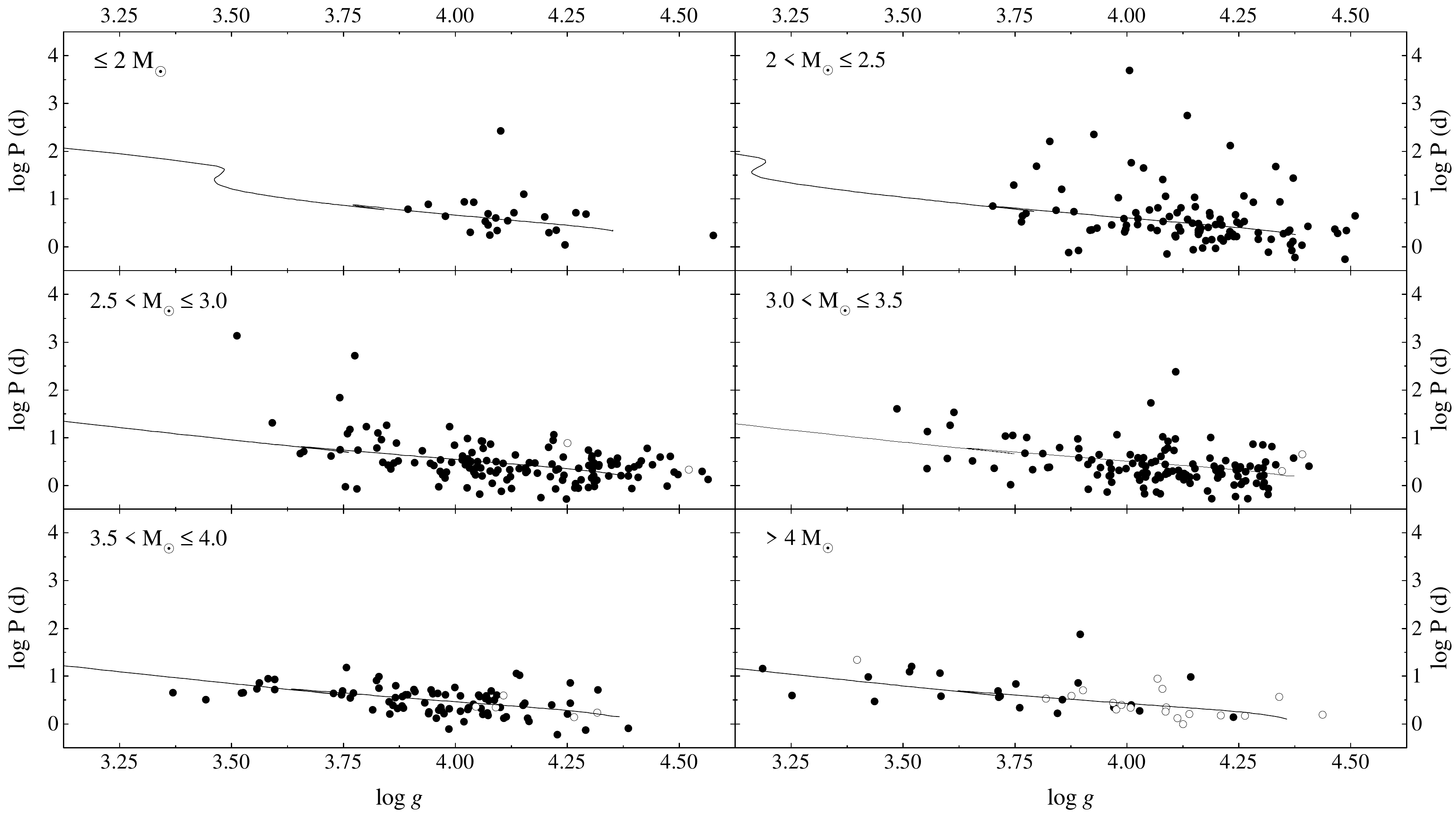}
    \caption{Rotation periods as a function of gravity (\logg) for the defined mass ranges. CP2 and CP4 stars are indicated by filled and open circles, respectively. The solid lines represent the rotation models by \citet{georgy13} for the masses and velocities listed in Table \ref{tab:mean_velo}.}
    \label{fig:rotevol}
\end{figure*}

Table \ref{tab:mean_velo} and Fig. \ref{fig:mass_velo} show the median rotational properties for the individual mass ranges. In order to reduce the influence of stellar evolution on $V_{\rm eq}$, we excluded potentially evolved stars by imposing the restriction \logg\ $> 4.0$, but for the calculation of the $\upsilon/\upsilon_{\rm crit}$ ratio, the complete sample was used. However, objects deviating by more than 2$\sigma$ from the initial mean values were excluded in the calculations.
Furthermore, we took care that stars with longer periods do not distort the results by a careful evaluation of the distribution of periods in the respective mass groups (cf. Fig. \ref{fig:rotevol}). If an outlying long-period object was found to exact a profound influence on the resulting fit, it was removed from consideration. In this way, a few mostly lower mass stars with periods longer than 25 d were removed. For comparison, we have also provided the values of $V_{\rm eq}$ that were derived from the observed $\upsilon/\upsilon_{\rm crit}$ ratio and the model of the ZAMS by \citet{ekstrom12}. These agree well within the errors, although values from the latter source are all higher by a small amount owing to evolutionary effects. As can be seen in Table \ref{tab:mean_velo} and Fig. \ref{fig:mass_velo}, the median absolute deviations are quite large and generally exceed the individual errors of the derived equatorial velocities and velocity ratios. Thus, there is either an intrinsic spread of the velocities or other properties (most likely the magnetic field strength) might introduce an additional broadening. However, a clear increase of the velocities with mass is noticeable. If we exclude the highest mass group, which covers by far the largest range of masses and therefore might introduce a distortion, we derive the linear dependence 
\begin{equation}
\label{vcritmass}
\upsilon/\upsilon_{\rm crit} = 0.006(5)+0.040(2)M/\msun
\end{equation}
with a correlation coefficient $R=1.0$. We note that the inclusion of the highest mass group does not alter this result significantly and still provides $R=0.99$.

Fig. \ref{fig:rotevol} shows the evolution of the rotational periods as a function of \logg\ for the individual mass groups. Furthermore, as comparison we include the rotation models by \citet{georgy13}, which were interpolated to match the observed velocity ratios for the respective masses as listed in Table \ref{tab:mean_velo}. A very good agreement with the models that follow conservation of AM was noticed for all mass ranges. Thus, strong additional magnetic braking along the MS can be ruled out, at least for a ``typical'' mCP star. This confirms the results of previous studies, such as \citet{north98} or \citet{stepien98}, who found evidence for the conservation of AM during the stellar evolution of mCP stars. However, as briefly mentioned in Section \ref{rotation_mCP}, a change of the rotation period has been observed in a few mCP stars.

Most of the stars with the longest periods are found in the mass range $2.0\,\msun < M \leq 2.5\,\msun$, which is also noticeable in the work of \citet{kochukhov06}. \citet[][]{Preston1970} presented 25 CP2 and 10 CP3 stars with \vsini\ values lower than 10\kms, for which he inferred a rotational period longer than 16\,d. We have to emphasize that the 25 CP2 stars are all of SrCrEu type. \citet[][]{wolff75} later confirmed these results, adding that no Si star with a rotational period longer than 20\,d is known. However, in Fig. \ref{fig:rotevol}, few higher-mass stars with longer periods are included. As was shown by \citet[][]{Deutsch1967}, a Maxwellian distribution satisfactorily describes the distribution of rotational velocities of normal upper MS stars. With the availability of extended data sets with higher precision, a bimodal distribution was found \citep[][]{abt95} and an excess of slow rotators was established (cf. also Section \ref{rotation_mCP}). Both subsamples follow individual Maxwellian distributions with a mode of $\sim$50 and $\sim$200\kms \citep[][]{zorec12}, respectively. 

\citet[][]{Dworetsky1974} investigated the rotational velocities of A0 stars in a statistical sense. The final list includes 215 stars. For the statistical analysis, the \vsini\ values were binned to 40\kms. Within this sample, he found that 8\% of all 
stars are intrinsically slowly rotating stars. This number corresponds to a overrepresentation factor of 2.5 (see Fig. 3 therein).
\citet{Hubrig2000} investigated a sample of 160 A-type stars for which precise $Hipparcos$ parallaxes and \vsini\ values are known. 
They used a binning size of 30\kms. It is concluded that they do not find any excess of slow rotators. However, the distributions of the rotational velocities as well as \vsini\ values (Figs 3--6 therein), divided for apparently young and old stars, clearly show a significant excess of slowly rotating stars compared to expectations from the distribution of the true rotational velocities. Their
conclusion, on the other hand, is based on the fit of a Maxwellian distribution of the observed rotational velocities not corrected for
the orientation of the rotational axes. Because they have not published the used \vsini\ values, we deduced from the combined numbers of Figs. 5 and 6, an over-representation factor of about 5.

Using both, the large starting sample of about 1300 stars and the final working sample, and the expected frequency of long-period mCP stars ($>$\,160\,d) by \citet{wolff75}, we obtain an excess factor of about 10. This factor increases to about 30 and 40 if we consider stars with spectral types later than about A0. However, we note that the samples cannot be considered statistically representative, and as already mentioned in Section \ref{rotation_mCP}, the estimation of the actual excess of very slow rotators is not a trivial task and beyond the scope of this paper.

Within this context, the question arises how many stars with periods of years or even decades have not yet been detected. One of the first attempts to measure periods of this order was undertaken in the framework of the Long-Term Photometry of Variables programme at the European Southern Observatory \citep[][]{Sterken1995}. During 12 yr (1982--1994), high precision photoelectric observations were performed, which resulted in the detection of rotational periods for many CP stars. More recent surveys based on CCD observations, such as ASAS-3 (2000 -- 2009) and SuperWASP (since 2004), suffer from instrumental and reduction-based long-term trends that are difficult to distinguish from true intrinsic stellar variability \citep[][]{bernhard15b,huemmerich16}. Similar problems arise when data sets obtained with different instruments and passbands are merged \citep[][]{mikulasek10}. One has to keep in mind that the amplitudes of variability observed in mCP stars generally do not exceed $\sim$0.05\,mag \citep[][]{Manfroid1994}, which implies that a scatter of less than 0.01\,mag has to be maintained over a period of years. To our knowledge, in the last decade, no efforts were made to merge and analyse archival and new photometric observations in order to search for periods on time-scales of years. An investigation into this matter is sure to result in the detection of new mCP stars with very long periods, which would further increase the observed discrepancy among the slowly rotating objects.

About 60 mCP stars of the cool SrCrEu type also exhibit photometric variability in the period range of 5--25 min (high-overtone, low-degree, and non-radial pulsation modes). These stars are known as rapidly oscillating Ap (roAp) stars \citep{kurtz82}, and several of the slowest rotating cool mCP stars (e.g. $\gamma$ Equulei) belong to this group. Generally, only a very narrow astrophysical parameter and atmosphere regime seems to actually allow the excitation of roAp pulsations, which explains the rarity of these objects.
 \citet{kurtz82} introduced the oblique pulsator model, which assumes that the star pulsates about its magnetic axis. The observed frequency is then split in three components with amplitude ratios depending on the alignment of the magnetic and rotational axis. \citet{kochukhov05} investigated the effect of the rotational period on the pulsational amplitude. He found that slow rotators tend to have larger amplitudes, depending on the effective temperature. However, a detailed theoretical investigation into this matter has yet to be performed.

\section{Influence of magnetic field strength on stellar rotation}

Much effort has been spent in the investigation of the causes behind the observed spin-down in mCP stars. The ubiquitous presence of magnetic fields in these objects, which can potentially affect their rotational properties, offers a natural starting point for investigations \citep{mathys15}.

The currently favoured theory supposes a fossil origin of the field, i.e. it is supposed to be a relic of the interstellar magnetic field that has been ``frozen'' into the stellar plasma. Several correlations between the magnetic and rotational properties of mCP stars have been proposed. \citet{kochukhov06} found evidence that massive stars ($M$\,$\geq$\,3.0\,\msun) are intrinsically more magnetic than their low-mass analogues. Furthermore, a correlation between magnetic field and rotation period for intermediate-mass stars (2.0\,\msun$\leq$\,$M$\,$\leq$\,3.0\,\msun) was suggested. \citet{hubrig07} also postulated a preference for stronger magnetic fields to occur in younger and more massive stars. \citet{mathys16b} found that the strongest magnetic fields are solely found in stars with rotation periods shorter than $\sim$150 d. Furthermore, it seems to be established that stars with $P$\textsubscript{rot}\,<\,100 d and $P$\textsubscript{rot}\,>\,1000 d exhibit large values of the angle between the magnetic and rotation axes ($\beta$). However, more long-term data are needed to corroborate most of the above mentioned assumptions \citep{mathys16b}, and it is not clear in what way they are connected to the actual processes at work in the spin-down of mCP stars.

As has been outlined in Section \ref{rotation_mCP}, the currently favoured theory evoked to explain the observed differences in rotational velocities among early MS and mCP stars is AM loss during the PMS phase of stellar evolution. The most detailed investigation into the underlying causes is the seminal work by \citet{stepien00}, who discussed the early evolution of mCP stars and explained the observed slow rotation as being due to the interaction of the stellar magnetic field with the circumstellar environment during the PMS phase.

However, some mCP stars rotate with periods in the range of years to centuries. According to the scenario invoked by \citet{stepien00}, these very slow rotation periods can be achieved only in the case of lower mass stars with correspondingly long PMS lifetimes. After an early disappearance of the circumstellar disc, strong magnetized winds can slow down rotation during the long time of approach to the ZAMS, resulting in periods up to $\sim$100 years.

Observational evidence seems to confirm that extremely slowly rotating mCP stars are found among lower-mass stars with strong magnetic fields \citep{stepien00}. Evidence has been mounting, though, that other mechanisms might also play a part in the spin-down of these extreme objects as stars with the longest periods do not necessarily show also the very strongest magnetic fields \citep{mathys16b}.

In summary, the presence of a primordial magnetic field of moderate strength is able to explain the observed mean rotation rates of chemically normal early MS stars and mCP stars \citep{stepien98}. The resulting ZAMS period of mCP stars is determined by the details of the complex interaction between the stellar magnetic field and the circumstellar environment during the PMS phase \citep{stepien00}.

For the study of the dependence of rotation on magnetic field strength, we intended to use magnetic field indicators that represent the surface field strength best. The mean magnetic field modulus $\langle B \rangle_{\rm av}$ certainly provides the best measure of the actual field strength, but resolved magnetically-split lines can generally be measured only in (very) slow rotators with $\langle B \rangle_{\rm av}$ $\ga$ 2.7\,kG \citep{mathys16b}. \citet{mathys16b} derived and compiled corresponding data for some dozen stars. On the other hand, the longitudinal magnetic field is by far the most often measured parameter \citep[e.g.][]{bychkov09}, but it is also prone to the largest variations with phase. Therefore, previous works \citep[such as][]{kochukhov06} have adopted the average quadratic longitudinal field $\langle  \overline{B_\ell} \rangle $ \citep{borra83} in order to reduce the influence of the variability to some extent. However, statistical conclusions based on this value are limited by the available phase coverage. \citet{bychkov05} and \citet{auriere07} analysed rotational phase curves of longitudinal magnetic field measurements. Based on the average field $B_{0}$ and half of the amplitude $B_{1}$, one obtains $B_{\ell}^{\rm max} = \left|B_{0}\right| + B_{1}$, and, multiplied by a factor of 3.3, a lower limit of the surface dipole component of the magnetic field \citep[see discussion by][]{auriere07}.

For our sample, we have finally adopted measurements of the magnetic field strength in the following order: \citet{mathys16b}, \citet{auriere07}, \citet{bychkov05}. For the remaining stars, we have used $\langle  \overline{B_\ell} \rangle $ from the list by \citet{bychkov09} as the lowest quality indicator of the magnetic field strength. From this source, only values based on at least three measurements were considered which were multiplied by a factor of 3.3; we emphasize that these values still represent a lower limit of the actual magnetic field strength. Employing these criteria, we have compiled corresponding data for 160 stars in total. In order to compare stars in different evolutionary stages, a correction following conservation of magnetic flux $(B\,R^2)$ is needed. We have therefore derived the radius ratio $R/R_{Z}$ using the observed radius $R/{\rm R}_\odot$ and the corresponding radius at the ZAMS ($R_{Z}$) based on the models by \citet{ekstrom12}. We note that we have adopted $R/R_{Z}=1$ for the few stars located below the ZAMS (see Fig. \ref{fig:hrd}).

\begin{figure}
	\includegraphics[width=\columnwidth]{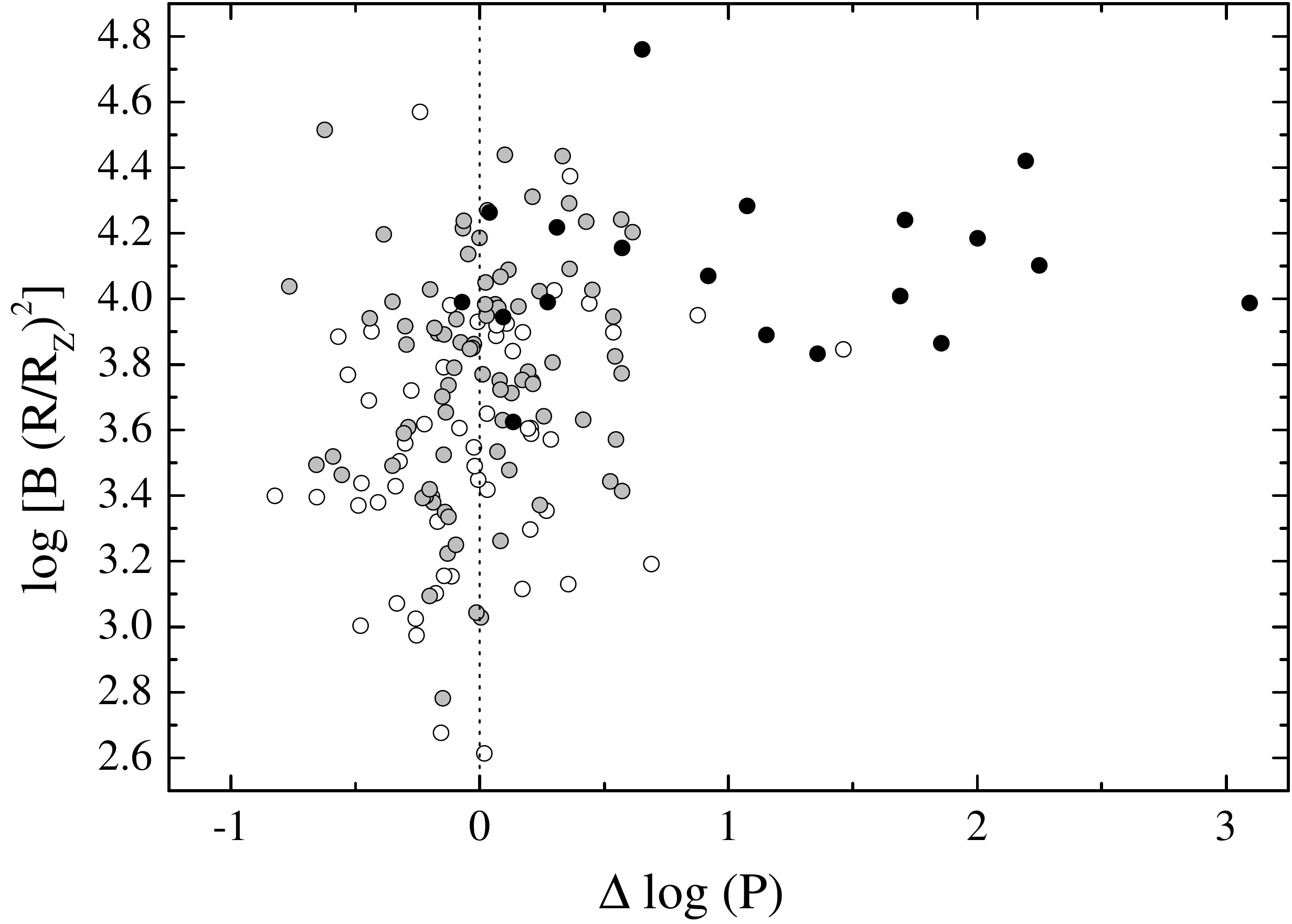}
    \caption{Magnetic field strength as a function of the rotation period, the latter corrected by evolutionary and mass effects using the adopted models (i.e. the differences between observed and model value as seen in Fig. \ref{fig:rotevol}). Black circles represent measurements of the mean magnetic field modulus, grey circles estimates based on rotational phase curves of the longitudinal magnetic field and open circles the average quadratic longitudinal field. We note that the last two represent lower limits of the actual magnetic field strength. The vertical dotted line separates stars that rotate faster (left-hand side) and slower (right-hand side) than predicted by conservation of AM.}
    \label{fig:period_mag}
\end{figure}

Fig. \ref{fig:period_mag} shows the magnetic field strength as a function of rotation period; the latter has been corrected for effects of stellar evolution and mass. Thus, we have used $\Delta \log (P)$ as the difference between the observed period and the corresponding model value (i.e. the residuals from Fig. \ref{fig:rotevol}). Some conclusions can be already drawn from this figure. All stars with the longest periods have strong magnetic fields, but there are much stronger fields among objects with shorter periods, confirming the findings of e.g. \citet{mathys16b}. We note that the star with the strongest magnetic field is HD~215441 (also known as Babcock's star), which the author of the last reference considers as a non-representative member of the mCP star group. The comparison of the magnetic field strength of faster and slower rotating stars (employing the vertical line in Fig. \ref{fig:period_mag} as threshold) shows that the slower rotating objects have magnetic fields about a factor of two stronger than encountered in the faster-rotating stars (median values of $\sim$\,8 and $\sim$\,4\,kG, respectively). We have applied a $t$ test, from which we derived a significance of $\gg$ 95\,\%.

However, there still is the need to explain the relationship between the (very) long-period stars with somewhat lower magnetic field strengths and the faster rotating stars with very strong magnetic fields (such as HD~215441). The question arises whether we actually have to consider additional factors beside the magnetic field or whether HD~215441 is indeed an atypical star, at least in this respect? So far, we have not considered a possible dependence on mass, and an examination just in the period domain might be delusive. Let us compare the properties of the star with the longest period in the sample (HD~110066) to Babcock's star (see Table \ref{table_compstars}). Clearly, the difference in the magnetic field strength cannot be explained by conservation of magnetic flux, because both objects show comparable $R/R_{Z}$ values and, owing to the very long period, the rotation rate $\upsilon/\upsilon_{\rm crit}$ of HD~110066 is almost zero ($\sim 10^{-4} \pm 15\,\%$). However, the masses of these two objects differ significantly (cf. Table  \ref{table_compstars}). As shown in Fig. \ref{fig:mass_velo} and Equation \ref{vcritmass}, the typical rotation rates of mCP stars with these masses are 0.10 and 0.17. The difference is larger than we find from the observed rotation rates of the two stars mentioned above. Thus, if we take the mass dependency into account, Babcock's star (as the more massive one) was actually braked more efficiently than HD~110066 owing to its extraordinary strong magnetic field. This shows that a direct comparison might be misleading and that we need to investigate rotation rates that are corrected by the mass. 

\begin{table}
\caption{Comparison of the mCP stars HD~110066 and HD~215441.}
\label{table_compstars}
\begin{center}
\begin{tabular}{lll}
\hline 
   & HD~110066 & HD~215441 \\
\hline
Period & 4900\,d & 9.5\,d \\
$\langle B \rangle_{\rm av}$ & 4.1\,kG & 33.6\,kG \\
$M/\msun$ & 2.4 $\pm$ 0.1 & 4.2 $\pm$ 0.3 \\
$R/R_{Z}$ & 1.5 $\pm$ 0.2 & 1.3 $\pm$ 0.3 \\
$\upsilon/\upsilon_{\rm crit}$ & 0.00 $\pm$ 0.00 & 0.04 $\pm$ 0.01 \\
\hline
\end{tabular}
\end{center}
\end{table}

\begin{figure}
	\includegraphics[width=\columnwidth]{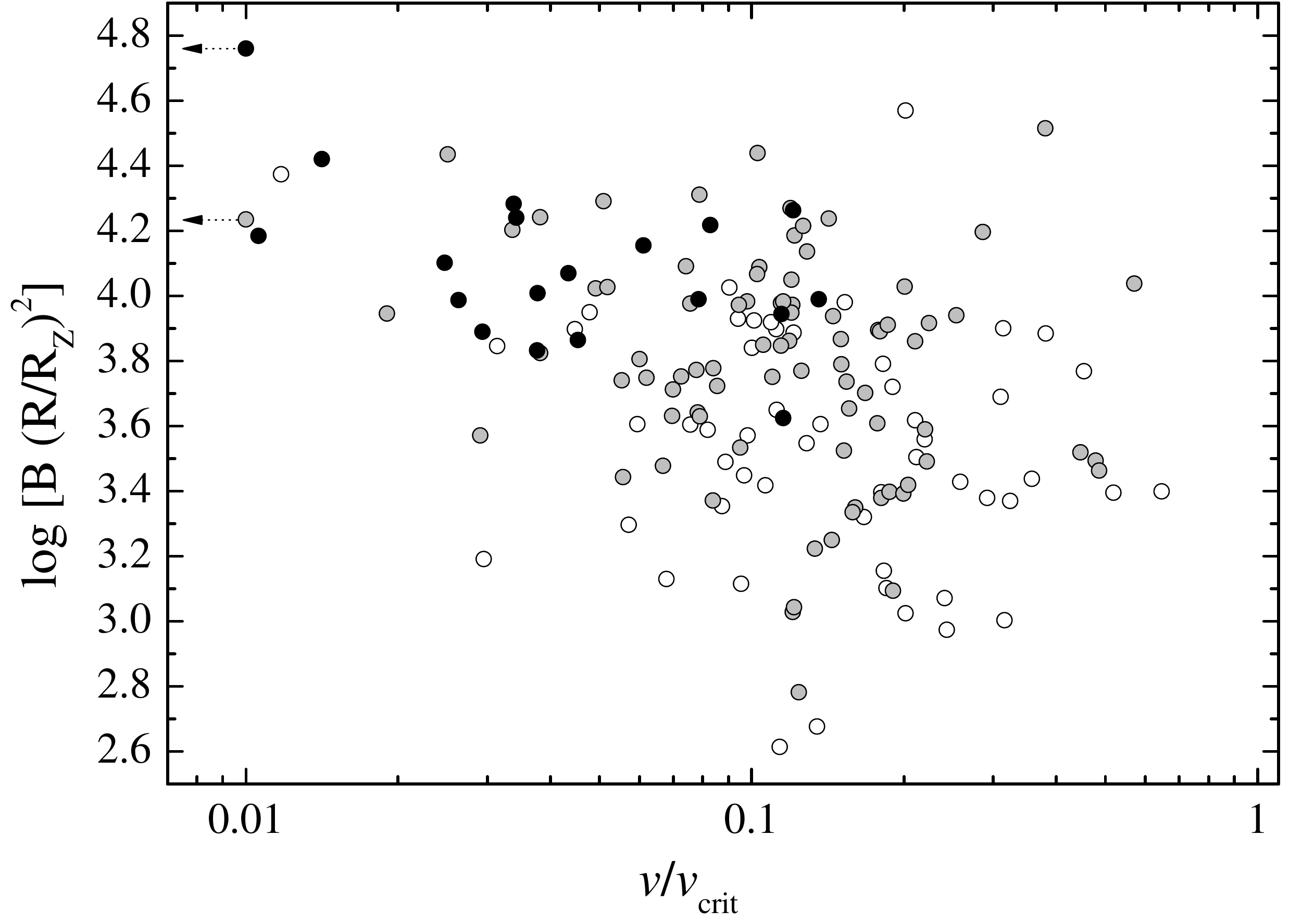}
    \caption{Magnetic field strength as a function of the rotation rate $\upsilon/\upsilon_{\rm crit}$, the latter normalized to the rotation rate of a 3\,\msun\ mCP star. Two stars (HD~215441 and HD~175362) show a negative $\upsilon/\upsilon_{\rm crit}$ after this correction, we thus adopt 0.01 (the error) as possible upper limit. The symbols are the same as in Fig. \ref{fig:period_mag}, and we note again that the grey and open symbols represent lower limits of the actual magnetic field strength.}
    \label{fig:vvcrit_mag}
\end{figure}

We have used this information to show our sample in the $\upsilon/\upsilon_{\rm crit}$ domain, normalized to the rotation rate of a 3\,\msun\ mCP star using the slope of Equation \ref{vcritmass}. Fig. \ref{fig:vvcrit_mag} clearly indicates that after this normalisation process, the stars with the strongest magnetic fields are in general also the slowest rotators. Note that we needed to exclude several of the highest mass stars ($>$ 5\,\msun) because of the unknown continuation of the mass/velocity relation. This might have influenced the result for HD~175362 ($\sim$\,4.7\msun) in Fig. \ref{fig:vvcrit_mag}. This star and HD~215441 show a negative $\upsilon/\upsilon_{\rm crit}$ ratio after the normalisation process; we therefore adopt the error as upper limit.

We have searched our sample for objects that might be good candidates for exhibiting extraordinarily strong magnetic fields, which makes them ideal targets for follow-up observations. HD~102797, for example, boasts the same mass as Babcock's star, but shows a rotation period of 75\,d as listed in VSX (see also Fig. \ref{fig:rotevol}), thus about eight times longer than the `magnetic record holder'. The period of the 3.5\,\msun\ star HD~69544 (236.5\,d) was recently derived by \citet{bernhard15a}, but the object is otherwise little studied.

Fig. \ref{fig:vvcrit_mag} indicates some faster rotating stars with strong magnetic fields that we want to discuss briefly. HD~184905 is the star with the strongest magnetic field in the lower quality $\langle  \overline{B_\ell} \rangle $ sample, though the reference of the measurement is a private communication from 2002 \citep[see][]{bychkov09} and has apparently still not been published yet. For the object HD~111133, the strongest magnetic star in the sample with magnetic phase curves, there might be a problem with identifying the real rotation period, as discussed by \citet{wraight12}. If we adopt the period of 16.3\,d as used by \citet{bychkov05} instead of 2.2\,d, we derive a normalized value of $\upsilon/\upsilon_{\rm crit}=0.05$.

We want to emphasize that our main intention was the re-evaluation of evolutionary changes of rotation based on a much larger sample of mCP stars than had been available to the authors of previous studies. Thus, the sample is not best suited for a more detailed investigation of the magnetic properties. For example, the sample includes only a small fraction of the stars listed by \citet{mathys16b}. This might be owed to the selection of references our compilation of periods has been based on, the exclusion of spectroscopic binary systems, but also to the adopted quality criteria or missing information on the parallax. The last two issues will certainly be improved with the advent of the next $Gaia$ data releases. Therefore, in a follow-up work, we intend to reverse our current approach by starting from a list of stars with well-known magnetic field strengths, and to compile (or derive) all other relevant information.

\section{Summary and conclusions}

We have presented an analysis of the rotational properties of magnetic CP stars based on an unprecedentedly large sample consisting of more than 500 objects. Our investigation was made possible by several recent works that investigated the rotational periods of mCP stars and, in particular, the availability of the first data release of the $Gaia$ satellite mission. We have derived astrophysical parameters for our programme stars such as effective temperature, luminosity and finally mass, equatorial velocity and the rotation rate $\upsilon/\upsilon_{\rm crit}$. Our sample covers the well-known mass range of mCP stars ($1.5\,\msun \la M \la 7\,\msun$), and the large number of objects allows to define narrower mass groups than have been employed in previous investigations.

We have confirmed that the evolution of rotational periods among mCP stars agrees very well with the assumption of conservation of AM. No evidence for significant additional magnetic braking along the MS was found. For this analysis, we have employed recent stellar evolutionary models that take into account rotation and conservation of AM, and we have briefly discussed the differences between the available models. We note that our sample is also well-suited as test-bed for the investigation of future evolutionary models that take into account magnetic fields -- a certainly challenging inclusion.

Furthermore, as generally expected, we have confirmed that the inclination angles of the rotational axes are randomly distributed, although there is an apparent excess of fast rotators sharing comparable inclination angles. The reason for this remains unclear yet. The group of very fast rotators is small, but by no means negligible. We identified 14 objects that show a rotation rate of $\upsilon/\upsilon_{\rm crit} \geq 0.5$, corresponding to equatorial velocities $V_{\rm eq} \sim 200$\,\kms. Theoretic models still need to explain the interplay between diffusion and mixing processes in the presence of strong ($\sim$ 1.5\,kG) stable magnetic fields at these high rotational velocities.

Stars on the other side of the velocity distribution, the very slow rotating mCP objects, have obtained much more attention so far. Our sample includes stars up to rotation periods of 4900 d, but for some objects even much longer periods were inferred in the literature. All of these objects show very strong magnetic fields, although by far not the strongest ones. Previous studies conclude that the strongest magnetic stars have periods of less than about 150 d, and that some of them (in particular Babcock's star) should be considered as atypical representatives of the mCP group. However, as has been shown in this work, an investigation that only focuses on the period domain can be delusive.

To our knowledge, we have derived for the first time the relationship between mass and rotation rate of mCP stars. Nevertheless, we note that a larger sample of massive mCP stars is needed to trace this relation beyond $\sim$\,5\msun. The dependence between mass and rotation was employed to link these two parameters to the magnetic field strength. Babcock's star is more massive than the very long-period objects, and its rotation rate, corrected for mass, is therefore actually lower, and it was braked more efficiently owing to its very strong magnetic field. Thus, an anticorrelation between magnetic field and rotational velocity is observed: the stronger the magnetic field, the slower the rotation (in a relative sense). In addition, we have presented several good candidates for exhibiting extraordinarily strong magnetic fields that might even exceed the one observed in Babcock's star.

\section*{Acknowledgements}
MN acknowledges the support by the grant 14-26115P of the Czech Science Foundation. This work was also supported by the grant 7AMB17AT030 (M\v{S}MT). We made use of the SIMBAD data base and the VizieR catalogue access tool operated at CDS, Strasbourg, France, and VOSA, developed under the Spanish Virtual Observatory project supported from the Spanish MICINN through grant AyA2011-24052. The work has also made use of data from the European Space Agency (ESA) mission Gaia (http://www.cosmos.esa.int/gaia), processed by the Gaia Data Processing and Analysis Consortium (DPAC, http://www.cosmos.esa.int/web/gaia/dpac/consortium). Funding for the DPAC has been provided by national institutions, in particular the institutions participating in the Gaia Multilateral Agreement. Furthermore, we acknowledge the use of data products from the Wide-field Infrared Survey Explorer, which is a joint project of the University of California, Los Angeles, and the Jet Propulsion Laboratory/California Institute of Technology, funded by the National Aeronautics and Space Administration. We thank Gautier Mathys for his helpful comments and suggestions that have greatly improved the manuscript.




\bibliographystyle{mnras}
\bibliography{cprot_bib} 




%
%


\bsp	
\label{lastpage}
\end{document}